\documentclass[usenatbib]{mn2e}
\usepackage{epsfig}
\begin{document}

\title[Variable stars in the core of M3]{Variable stars in the core of the globular cluster M3}

\author[Jay Strader, Henry O. Everitt, and Stephen Danford]{Jay Strader$^{1,2}$,
Henry O. Everitt$^{1,3}$, Stephen Danford$^4$\\ $^1$Department of Physics, Duke University, Box 90305, Durham, NC 27708\\
$^2$UCO/Lick Observatory, University of California, Santa Cruz, CA, 95064, strader@ucolick.org (current address)\\
$^3$U.S. Army Research Office, Box 12211, RTP, NC 27709, everitt@aro.arl.army.mil\\
$^4$Department of Physics \& Astronomy, University of North Carolina--Greensboro, P.O. Box 26170, Greensboro, NC, 27402, danford@uncg.edu}

\maketitle
\begin{abstract}
We present the results of a survey for variable stars in the core of the globular cluster M3. Our findings include the discovery of
eleven new or suspected variables, including a possible W Vir, and the first period determinations for thirteen
previously known variables.
\end{abstract}

\section{Introduction}

M3 is the most variable-rich galactic globular cluster, and the majority of these variable stars
are RR Lyraes (RRLs). Although the outer portions of the cluster were long ago surveyed for variables (e.g. \citealp{b13,s73}), the
identification and classification of RRLs in the core has proven much more difficult. Many variables in the core, though identified as such,
have poor or undetermined periods and light curves. The advent of image subtraction techniques for differential photometry, particularly the recent
work of \citet{a00}, has made systematic searches for RRLs in the cores of globular clusters with modest (i.e. $\leq$ 1-m) instruments possible. Such studies will
help to elucidate the radial distribution of the different classes of RRLs. In addition, the wide variety of pulsation behavior of RRLs serves as a testbed
for theories of stellar evolution, and regular monitoring of possible period changes in RR Lyraes may help to shed light upon such phenomena
as the Blazhko and Oosterhoff effects \citep{b07,o39}.

\section{Observations and Reductions}

\subsection{Observations}

All images were taken at the Three College Observatory (TCO), located in Snow Camp, NC, between Durham and Greensboro. Built in 1981 and operated by the
University of North Carolina-Greensboro, TCO houses a 0.81-m f/13.5 Ritchey-Chretien telescope. It is equipped with a 1024 $\times$ 1024 Texas Instruments 215 charge-coupled
device (CCD), binned to 512 $\times$ 512. The pixel scale is 0.46\arcsec, with a 3\arcmin51\arcsec $\times$ 3\arcmin51\arcsec field of view.

Observations were made on fifteen nights from March to July 2001. Seeing ranged from 1.5 to 2.0 arcsec. Images were taken in the \textit{V} band, and
exposure times were generally 160 s since longer exposures saturated the central part of the core. Eleven of the fifteen nights produced at least one image suitable for differential photometry. In total, 61 images were of sufficient quality to be used in subsequent analysis.
All images were taken with the CCD set to the same field.

\subsection{Reductions}

The Image Reduction and Analysis Facility ({\sevensize IRAF}) was used for the standard data reduction. All frames were trimmed, debiased, and flat-fielded using dome flats.
Since all analysis was performed using only differential photometry, no standard stars were observed to calculate coefficients for magnitude transformation equations.

\subsection{Image Subtraction}

The reduced images were processed using the image subtraction method (ISM) of \citet{al98}, as implemented in the package {\sevensize{ISIS 2.1}}, obtained through private communication with C. Alard.
Five of the images with the best seeing were combined to serve as the reference frame, which is shown in Figure 1. All images were registered to the reference frame, and then for each image,
a third-order convolution kernel was formed. The convolved reference frame was then subtracted from each individual frame. Then, for each pixel, a time series of
these residual fluxes was created. Data points more than three standard deviations from the median were removed from the time series to account for contaminating events, such as cosmic rays
or severe atmospheric disturbances. The median of this corrected set of data became a robust statistic for the variability of that pixel. The composite of all of these pixel statistics,
essentially a stacked residual image, was denoted \textit{var.fits}.

On \textit{var.fits}, the only stars visible were variable stars--primarily RR Lyraes. A routine was then run
to search for objects in \textit{var.fits}. To avoid missing variables in the crowded core, the detection limit was set to only one standard deviation, so that the program
found all objects whose `brightness' was more than 1$\sigma$ above the local background.

Due to this low threshold, a substantial number of detected objects were `false positives' in
that they did not represent actual variable objects. The majority of these detections
were associated with the extremely noisy edges of the \textit{var.fits} file, which are an
artifact of the ISM. These clear false positives were removed manually. A smaller number of
objects were later found to have constant light curves, and were then removed from the data sample.
One exception is the star v224, which is discussed in $\S 4.2.$

\subsection{Photometry}

For each of the remaining candidate variable stars, an unphased light curve was produced.
The {\sevensize{ISIS}} built-in routine for PSF photometry calculated the differential
flux for each of the residual images by convolving the reference frame PSF to the individual
frame using the previously calculated local kernel.

For stars $\ga$ 40 arcsec from the core, the internal dispersion of our photometry is
$\sim$ 3-6 per cent of the differential flux. A direct conversion into instrumental magnitudes
gives an upper limit to the internal error as 0.03 - 0.06 mag for this group of stars.
For stars closer to the crowded core, the errors rise to typical values of 8-10 per cent
(0.08 - 0.10 mag). Several stars within 25 arcsec of the core have
dispersions as high as 20 per cent (0.20 mag). It should, however, be kept in mind that
the \textit{absolute} fluxes of these stars are certainly greater than the differential fluxes
calculated, and thus the errors given here represent generous upper limits.

\section{Analysis}

\subsection{Period Searching}

To search for the periods of the suspected variables, Lomb-Scargle periodogram \citep{s82} and phase dispersion minimization (PDM) \citep{s78} methods were used,
as implemented in the R. Barbera's {\sevensize{AVE}} program\footnote{Version 2.51 may be found at http://www.astrogea.org/soft/ ave/aveint.htm}.
For each star, both methods were used to search for the best-fit period as described in this section. In nearly all cases, the resulting periods agreed to the fourth or fifth decimal place.
When discrepancies existed, visual inspection of the resulting light curves was used to choose the period that minimized the scatter in the light curve.

Since the focus of the search was on RRLs, the period range 0.2-0.8 days was searched. In a few cases, when no accurate period could be found in that interval, the range was extended at both ends to allow for the rare possibility of a Cepheid, RGB, or SX Phe star.

For the majority of the stars in the outer portions of the images, the PDM/periodogram (PDMP) analysis showed a single clear period and produced a relatively clean light curve when fit to the data.
For those stars closer to the core, however, the PDMP analysis did not always suggest a single period. In these cases, periods corresponding to all of the respective PDMP extrema were tried, and the
resulting light curves were compared. In many cases it was possible to eliminate other periods because they were incompatible with the type of the RRL, as indicated by the
shape of the light curve. If these methods still left multiple possible periods, the candidate light curves were inspected visually to ascertain the most accurate period.
For a small number of cases, no such determination could be made, and no period was assigned.

For a few stars, the noise in the light curve can be at least partially attributed to the presence of a nearby star (variable or not), whose light interferes with the signal from the variable star. These stars
have been marked as `merged' in Table 1.

\subsection{Identification}

After calculating preliminary periods for all of the detected variables in the images, RA/Dec coordinates were found for each star. To do this, 20 variables were identified
using the finder charts in \citet{c98} [C98]. Each variable had its identity confirmed using rough period matching from values given in the same source.
Since complete astrometry for all previously known variables is given in \citet{b00} [B00], it was possible to obtain accurate RA/Dec values for each of these known RRLs.
The {\sevensize IRAF} tasks \emph{ccmap} and \emph{wcsctran} were then used to compute transformation equations from physical (pixel) to world (RA/Dec) coordinates for the images,
and thus assign RA/Dec coordinates to each variable star in the sample. Since the accuracy of the B00 astrometry was $\leq$ 0.2 arcsec,
the process of matching variables in the sample to known RRLs was greatly facilitated.

For nearly all detected variables, the RA/Dec matches provided an accurate means of identifying RRLs in our sample. A second means of verification was period matching.
For each variable, the period was taken from the most recent source available, which was usually \citet{cc01} [CC01]. If the star was not included in the CC01 sample, an older period
determination could sometimes be found in catalogue of \citet{c01} [C01], an updated version of \citep{s73}.
\section{Results}

We identified 140 variable star candidates using the above procedures. Of these 140, 131 proved to be variable. Those variables previously known are listed in Table 1. The new and
published periods are compared, and comments are added where appropriate. Data for new and suspected variables is given in Table 2.

\subsection{Error Analysis}

The method used here calculates light curves using uncalibrated instrumental fluxes. This contrasts with most prior analyses of M3 RRL light curves which were plotted using standard apparent magnitudes. In this section, we address the self-consistency of this method and compare the results
to published photometry of RRLs.

\subsubsection{Self-Consistency}

To test the sensitivity of the period searching method to variations in the data, two stars were selected at random, one with a very clean light curve
and the other with a relatively noisy one. Five data points, equivalent to eight per cent of the total data set, were then randomly selected for each star and removed from the set of observations. The period
searching algorithm was then run as previously discussed.

v77 was chosen as the star with the clean light curve. Period analysis on both the full and reduced data sets gave the same period, 0.459419 d. The resulting light curve is slightly better
than that produced using the published value from CC01, 0.459350 d.

v214 was selected for its more scattered light curve. In this case, the best fit period changed from 0.539507 d to 0.539492 d. These are both close to the most recent published
period, 0.539493 d, taken from CC01. A visual inspection of the light curves for our two derived periods show no significant differences.

These results imply that our method is fairly robust: a slight pertubation of the data set does not have a substantial
impact on our derived periods to four decimal places.

\subsubsection{Comparison with Published Periods}

Another test of the accuracy of our method is how well it reproduces periods and light curves for well-studied stars. One difficulty is that we are studying primarily the inner core of M3, where the scatter in the data is inherently higher. However, many stars with previously
measured periods were included in our field of view.

For example, many variable stars in the uncrowded outer portions of M3 were studied here and by CC01.
For each of these variable stars, the light curves associated with the repsective periods were compared by PDMP analysis and by visual inspection. In many cases the period search algorithm was used on an interval that contained the CC01 period and
excluded ours to ascertain whether the CC01 period was a secondary extremum in the PDMP analysis of our data.

For each star studied by CC01 and included in our sample, we found that the period determined by our analysis gave a fit
that matched or exceeded the fit obtained using the CC01 period. In the latter cases, though the CC01 period might have a larger number of decimal places,
PDMP analysis and visual inspection suggested that our periods were a better match than the CC01 periods to our data. In total,
we found that 71 stars had period differences of less than 0.001 d between our period and the CC01 period.
Twenty-eight additional stars had period differences of 0.001 d or more.

Since M3 is an Oosterhoff I cluster, the expected period change rate ($\beta$) is relatively
low--CC01 found that typically, $\beta < 1 \textrm{ day Myr}^{-1}$.
Thus, given that the CC01 periods were determined using 1992-1997 data, it seems
unlikely that more than a few of these differences could be reasonably attributed to
period changes.

For seven of these 28 stars (v41, v148, v165, v170, v201, v209, v226) our periods agree with
the C98 or C01 periods to within 0.001 d.
Nine of the remaining 21 stars are listed as `blends' by CC01. Their color index
suggests the light of the variable star is mixed with that of a nearby non-variable star.
Two additional stars merged with other RRLs (v122 with v229, v241 with v262).
This merging may have led to a somewhat higher uncertainty in the period.
For these eleven stars, we could not make an unambiguous determination as to which period
gave a superior fit.

Seven of the remaining ten stars have noisy published light curves in CC01. Two of these
stars (v193, v235) had coverage too poor in our data set to make a definite determination
between our period and the CC01 period.
For each of the other five stars, we plotted the light curves for both periods, and present
them side by side for comparison in Figure 2. Visual inspection of these light curves indicates
that our periods give superior results for each of these five stars. Though three of these
five stars (v161, v175, v184) have periods included in C01, neither our periods nor those of
CC01 agree with the C01 periods to 0.001 d.

Only three of the original 28 stars with period discrepancies greater than 0.001 d have
both clean published light curves in CC01 and clean light curves when phased with our periods.
Further study of these variables (v149, v208, v219) is needed to determine the correct
periods.

\subsubsection{Accuracy}

These consistency tests and the fact that the data spanned between 200-400 cycles of most variables led us to conclude that our measured periods were accurate to approximately 0.0001 d. In some cases our periods
appear to be valid to at least 0.00001 d, and revised periods spanning many more cycles will be published in a future paper.
Therefore, all of the measured period values in Tables 1 and 2 are rounded to four decimal places. Published values are rounded to four places for comparison.

\subsection{Notes on Individual Variables}

Comments on some of the newly discovered, suspected, and previously known variable stars are
given below.  Light curves for the newly discovered variables, labeled S1-S11,
are presented in Figure 3, unless no period could be determined. For five of these stars
(S1, S2, S6, S9, S10), the quality of the photometry was too poor to ascertain
with certainty that the stars were variable, and thus we label these suspected variable stars.
Additional study of these stars is needed to constrain our identifications and period
determinations.

RR Lyrae types are identified using the
descriptive nomenclature of C01, based on pulsation modes. In this system, RR0 and RR1 denote fundamental and first overtone
pulsators, respectively, corresponding to Bailey types RRab and RRc. RR01 refers to
double-mode stars pulsating in both the fundamental and first overtone frequencies, commonly
called RRd stars.

\textbf{v29}: CC01 misidentifies this star as v155, an error present in a previous paper \citep{e94}.

\textbf{v122}: Although the light curve produced by our period, 0.5165 d, is noisy, that given by CC01, 0.5060 d, does not fit the data.
The discrepency is likely due to the merging of v122 with v229 on the images.

\textbf{v154}: Because our data covers only 12 full cycles of this star, it is likely that our period, 15.43 d, is less accurate than that published in C01 (15.28 d), although our results do agree with C01 to the twelve period ($\sim \textrm{eight per cent}$) accuracy our data allows.

\textbf{v224}: This star does not appear to be variable, but a definite determination could not be made using our data set.
			It is identified as variable X38 in \citet{k77}.

\textbf{v250}: There is no previous period determination for v250, for which there are two close candidate stars \citep{b00} in
			the WF/PC Hubble Space Telescope images of \citet{g94}. The confusion due to the companion resulted
			in an inconclusive PDMP, with several possible periods. Visual inspection of the resulting light curves suggests a period of 0.5586 d. 

\textbf{v273}: This star was identified as a long period variable by B00. Our analysis gives a period of 46.43 d, although this is poorly constrained because fewer than three cycles are covered by our observations. If this period is correct, the star could be a short period SR-type or a long period W Vir.

\textbf{S1}: This suspected variable has a large scatter in its light curve. Although a period of 0.3868 d best fits our data, this is unusually short for an RR0. A possible alternate period is 0.6310 d.

\textbf{S2}: The light curve of this suspected variable is very noisy, but apparently that of an RR1. There are no obvious period candidates besides 0.2980 d.

\textbf{S4}: Though 0.5075 d falls outside the normal range for RR1 variables, the light curve most closely resembles that of an RR1 and gives a clean fit to the data.
		Such a classification is not unprecedented in M3--the RR1 star v70 has a CC01 period of 0.4861 d. However, more data is needed to
		confirm this assignment.

\textbf{S6}: Although the period 0.6090 d gives the best fit, the light curve of this suspected variable is very noisy.

\textbf{S7}: No period in the normal RR Lyrae range (0.2 d - 0.8 d) fits our data. We find possible periods of 0.9883 d, 1.3270 d, and 4.0433 d, with the second of these
		producing the least scatter. The first two
		periods fall within the W Vir range, but since no magnitudes were calculated, we cannot confirm whether its period matches that predicted by the
		Population II Cepheid period-luminosity relationship. While it could be a star on the tip of the RGB, the best fit periods are far too short for the most
		common candidates (e.g. Mira, SR). Thus, until more observations are made, we label this a suspected W Vir, period 1.3270 d. 

\textbf{S8}: Our analysis suggests that this may be a long period variable, possibly an SRa or SRb. No unambiguous period measurement was possible over the limited time baseline of 119 d.

\textbf{S9}: For this suspected variable, the period 0.4261 d is favored by our analysis; however, large scatter near the minimum and poor coverage make this value uncertain. 

\textbf{S10}: While the best fit for our data, the period 0.4778 d nevertheless gives an extrememly noisy RR0 light curve for this suspected variable. 

\textbf{S11}: This star is likely an RR0 variable, but no clear period could be found.

\section{Conclusions}

We have conducted a survey of variable stars in the core of the globular cluster M3. We discovered eleven new or suspected variable stars, which include nine RR Lyraes,
one unidentified long period variable, and a suspected W Vir. For thirteen previously known variable stars, our period determinations are the first.
Additionally, we have updated period data for dozens of other poorly studied stars in the core. The success of this study is due largely to the effectiveness of the Alard ISM,
which enables systematic surveys of variable stars in the cores of galatic globular clusers using modest instruments.

\section*{Acknowledgments}

We thank C. Alard for helpful advice on the use of {\sevensize{ISIS}} 2.1, and appreciate R. Barbera's help in providing us with his {\sevensize{AVE}} program.
We also thank G. Wendt for the use of his period analysis software. G. Bakos kindly supplied finder charts that we used in our early analysis.
H. Smith, M. Corwin, and the referee provided useful suggestions which improved the manuscript.

\begin{center}
\begin{figure*}
\psfig{file=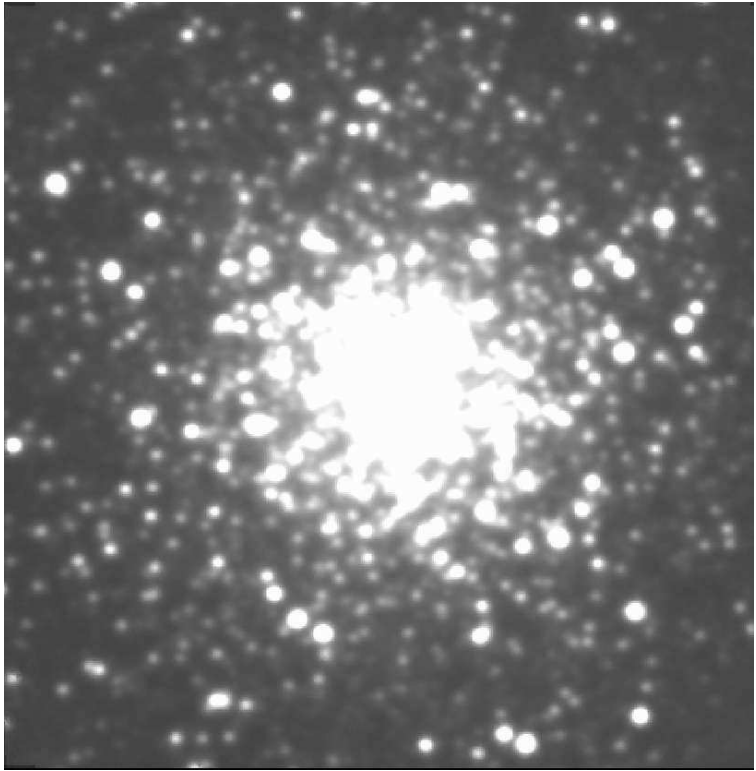,width=100mm}
\caption{The ISM reference image, which is a composite of images taken between March and May, 2001.}
\end{figure*}
\end{center}

\medskip

\begin{center}
\begin{figure*}
\psfig{file=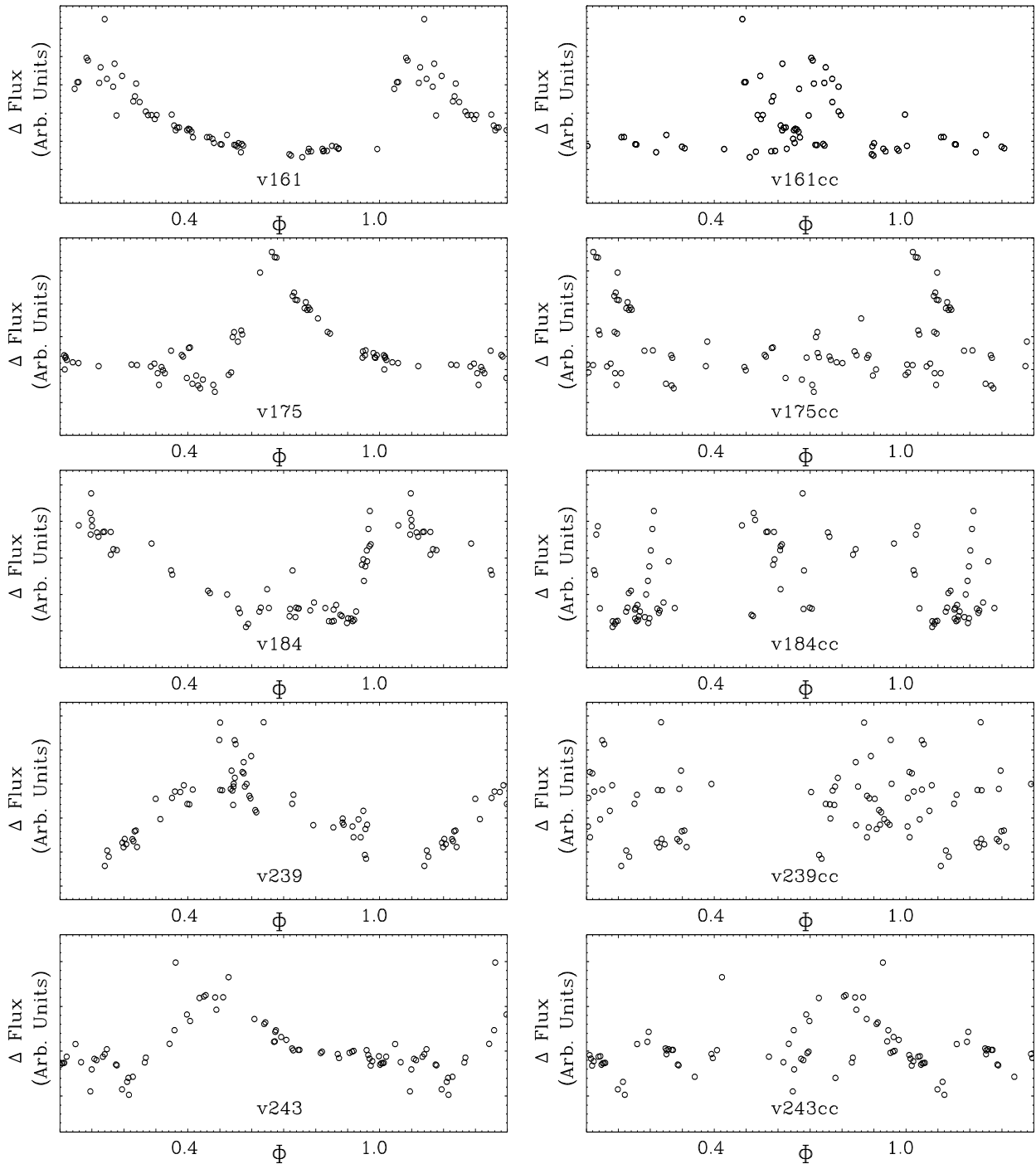,width=165mm}
\caption{Comparison light curves for five stars with disputed periods, as discussed in
$\S 4.1.2$. Light curves on the left are phased using the periods obtained through the PDMP
analysis of our data. Light curves on the right plot the same data using the CC01 periods.}
\end{figure*}
\end{center}

\begin{center}
\begin{figure*}
\psfig{file=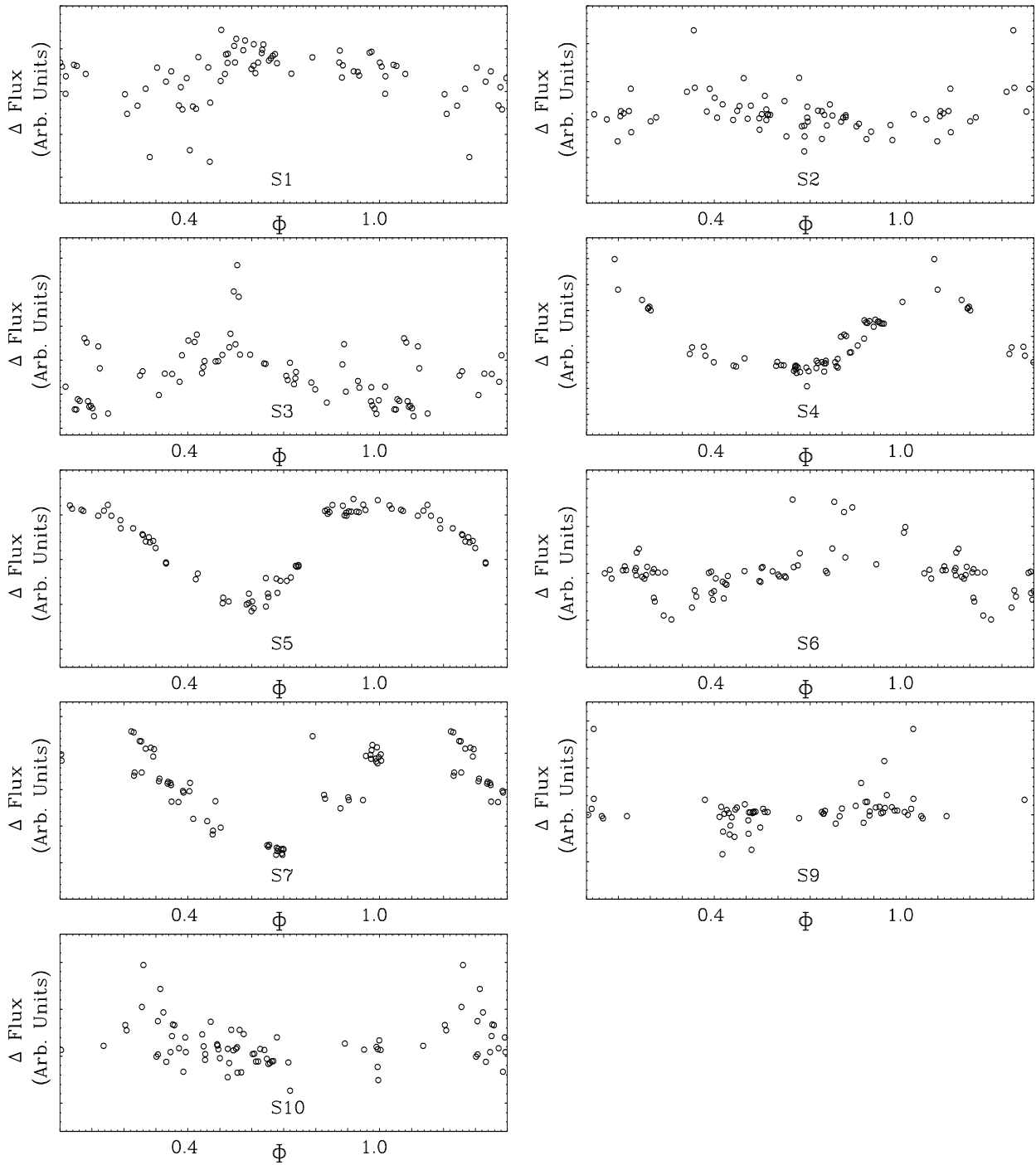,width=165mm}
\caption{Light curves for each of the newly discovered variable stars for which periods could be determined. (Table 2)}
\end{figure*}
\end{center}

\pagebreak

\begin{table*}
\begin{minipage}{160mm}
\caption{Data for previously identified variable stars in the core of M3.}
\begin{center}
\begin{tabular}{lccccclcccc}
\hline\hline
ID$^{\rm{a}}$ & Type$^{\rm{b}}$ & Period$^{\rm{c}}$ & Period (Pub.)$^{\rm{d}}$ & Comment$^{\rm{e}}$ & & ID$^{\rm{a}}$ & Type$^{\rm{b}}$ & Period$^{\rm{c}}$ & Period (Pub.)$^{\rm{d}}$ & Comment$^{\rm{e}}$\\
\hline
v3	&	RR0	&	0.5582	&	0.5582		&	 		&	& v173	&	RR0	&	0.6070	&	0.6070		& 			\\
v4	&	RR0	&	0.5849	&				&	n		&     & v174	&	RR0	&	0.5945	&	0.5945		&	n, pc		\\
v7	&	RR0	&	0.4974	&	0.4974		&	pc	 	&     & v175	&	RR0	&	0.5698	&	0.5658 		&	n, t		\\
v8	&	RR0	&	0.6371	&	0.6392		& 			&     & v176	&	RR0	&	0.5399	&	0.5406		&	n, pc		\\
v27	&	RR0	&	0.5790	&	0.5791		&	n		&     & v177	&	RR1	&	0.3481	&	0.3487		&	n		\\
v28	&	RR0	&	0.4706	&	0.4699		&	n		&     & v178	&	RR1	&	0.2670	&	0.2674		& 			\\
v29	&	RR0	&	0.4715	&	0.4711		& 	t		&     & v184	&	RR0	&	0.5313	&	0.5296		&	t 		\\
v30	&	RR0	&	0.5122	&	0.5121		&   			&     & v186	&	RR0	&	0.6633	&	0.6633		&	n		\\
v31	&	RR0	&	0.5807	&	0.5807		&  			&     & v187	&	RR0	&	0.4649	&	0.5841		&	n		\\
v32	&	RR0	&	0.4954	&	0.4954		& 	pc		&     & v188	&	RR1	&	0.2663	&	0.2663		& 	 		\\
v33	&	RR0	&	0.5252	&	0.5252		& 	n		&     & v189	&	RR0	&	0.6129	&	0.6187		&	 		\\
v41	&	RR0	&	0.4854	&	0.4866		& 	t		&     & v190	&	RR0	&	0.5228	&	0.5228		&	n	 	\\
v42	&	RR0	&	0.5901	&	0.5901		& 			&     & v191	&	RR0	&	0.5193	&	0.5192		&	n	 	\\
v43	&	RR0	&	0.5406	&	0.5405		& 			&     & v192	&	RR0	&	0.4972	&	0.4815		&	 	 	\\
v58	&	RR0	&	0.5171	&	0.5171		& 			&     & v193	&	RR0	&	0.7477	&	0.7328		&	pc, t	 	\\
v76	&	RR0	&	0.5018	&	0.5018		&	n		&     & v194	&	RR0	&	0.4890	&	0.4892		&	n, pc	 	\\
v77	&	RR0	&	0.4594	&	0.4594		&	t		&     & v195	&	RR0	&	0.6448	&	0.6439 		&	n		\\
v78	&	RR0	&	0.6120	&	0.6120		& 			&     & v197	&	RR0	&	0.4999	&	0.4999		&	pc		\\
v87	&	RR01	&	0.3575	&	0.3575		& 			&     & v200	&	RR0	&	0.5292	&	 			&	n, pc		\\
v88	&	RR1	&	0.2988	&	0.2985		&	n, pc 	&     & v201	&	RR0	&	0.5407	&	0.5396		& 	t		\\
v89	&	RR0	&	0.5485	&	0.5485		& 			&     & v207	&	RR1	&	0.4291	&	0.3449 		&	pc		\\
v101	&	RR0	&	0.6439	&	0.6439		& 	n, pc		&     & v208	&	RR1	&	0.3384	&	0.3373		&	n, t		\\
v109	&	RR0	&	0.5339	&	0.5339		& 	n		&     & v209	&	RR1	&	0.3483	&	0.3472		& 	t		\\
v110	&	RR0	&	0.5355	&	0.5355		&	n		&     & v212	&	RR0	&	0.5422	&	0.5422		&	n	 	\\
v111	&	RR0	&	0.5104	&	0.5102		& 			&     & v213	&	RR1	&	0.2994	&	0.2997 		&	n, pc		\\
v121	&	RR0	&	0.5352	&	0.5352		& 			&     & v214	&	RR0	&	0.5395	&	0.5395		&	n, t	 	\\
v122	&	RR0	&	0.5165	&	0.5060		&   m(v229), n, t &     & v215	&	RR0	&	0.5331	&	0.5331 		&	n		\\
v129	&	RR1	&	0.4056	&	0.4061		& 			&     & v216	&	RR1	&	0.3465	&	0.3465 		&	 	 	\\
v131	&	RR1	&	0.2977	&	0.2977		& 			&     & v218	&	RR0	&	0.5440	&	0.5431		& 	n	 	\\
v132	&	RR1	&	0.3398	&	0.3399		& 			&     & v219	&	RR0	&	0.6138	&	0.6114 		&	n, t	 	\\
v133	&	RR0	&	0.5508	&	0.5507		& 			&     & v220	&	RR0	&	0.5999	&	0.5959 		&	pc	 	\\
v134	&	RR0	&	0.6181	&	0.6181		& 	n		&     & v221	&	RR0	&	0.6098	&	0.3787 		&	pc	 	\\
v135	&	RR0	&	0.5684	&	0.5684		& 	n		&     & v222	&	RR0	&	0.5007	&	0.5007		&	n	 	\\
v136	&	RR0	&	0.6173	&	0.6172		& 			&     & v223	&	RR1	&	0.3293	&	0.3296 		&			\\
v137	&	RR0	&	0.5752	&	0.5752		& 			&     & v224	& 	 	&	 		&	 			&	nv		\\
v139	&	RR0	&	0.5601	&	0.5600		& 			&     & v226	&	RR0	&	0.4887	&	0.4877		&	n, t		\\
v142	&	RR0	&	0.5687	&	0.5686		& 	pc		&     & v229	&	RR0	&	0.6877	&	 			&	m(v122), n, t \\
v143	&	RR0	&	0.5965	&	0.5965		& 			&     & v234	&	RR0	&	0.5495	&	 			&	n, pc 	\\
v144	&	RR0	&	0.5968	&	0.5968		& 			&     & v235	&	RR0	&	0.7594	&	0.7617 		&	n, pc, t 	\\
v145	&	RR0	&	0.5145	&	0.5145		& 	n		&     & v239	&	RR0	&	0.6766	&	0.3334 		&	n, t  	\\
v146	&	RR0	&	0.5967	&	0.5967		& 	n		&     & v240	&	RR0	&	0.2760	&	0.2765 		&			\\
v147	&	RR1	&	0.3465	&	0.3465		& 			&     & v241	&	RR0	&	0.5450	&	0.5962 		&	m(v262), n, t \\
v148	&	RR0	&	0.4673	&	0.4660		& 	t		&     & v242	&	RR0	&	0.6513	&	 			&	n	  	\\
v149	&	RR0	&	0.5482	&	0.5500		& 	pc, t		&     & v243	&	RR0	&	0.6351	&	0.6322 		&	n, t 	    	\\
v150	&	RR0	&	0.5240	&	0.5239		& 			&     & v245	&	RR1	&	0.2840	&	 			&	 		\\
v151	&	RR0	&	0.5170	&	0.5178		& 			&     & v246	&	RR1	&	0.3392	&	0.3384 		&	n		\\
v152	&	RR1	&	0.3261	&	0.3261		&      	 	&     & v250	&	RR0	&	0.5586	&	 			&	m, n, t	\\
v154	&	W Vir	&	15.23		&	15.28			&	t		&     & v253 	&	RR1	&	0.3328	&	 			&	n, pc		\\
v156	&	RR0	&	0.5320	&	0.5320		& 			&     & v254	&	RR0	&	0.6056	&	 			&	 	   	\\
v157	&	RR0	&	0.5427	&	0.5419		&			&     & v256	&	RR1	&	0.3181	&	0.3156 		&	n		\\
v159	&	RR0	&	0.5337	&	0.5337		&	n		&     & v257	&	RR0	&	0.6019	&	 			&	n		\\
v160  &	RR0	&	0.6572	&	0.6573		&	pc		&     & v258	&	RR0	&	0.7134	&	0.7134 		&	n		\\
v161	&	RR0	&	0.5264	&	0.5144		& 	pc, t		&     & v259	&	RR0	&	0.5009	&	0.3335 		&	n, pc		\\
v165	&	RR0	&	0.4836	&	0.4850		& 	t		&     & v261	&	RR0	&	0.4447	&	0.4447 		&	 	   	\\
v166	&	RR01	&	0.4852	&	0.4852		& 			&     & v262	&	RR0	&	0.5647	&	 			&	m(v241), n, t \\
v167	&	RR0	&	0.6440	&	0.6440		& 	n		&     & v264	&	RR1	&	0.3563	&	 			&	n		\\
v168	&	RR1	&	0.2760	&	0.2764		& 			&     & v270	&	RR0	&	0.4938	&	0.6903		&	n, pc		\\
v170	&	RR1	&	0.4324	&	0.4357		& 	t		&     & v271	&	RR0	&	0.6329	&	0.6327		&	n		\\
v171	&	RR0	&	0.3032	&	0.3038		& 			&     & v273	&	lp	&	46.43		&	 			&	pc, t		\\
v172	&	RR0	&	0.5423	&	0.5423		&	n       	&     & 		&		&			&				&			\\
\hline

\end{tabular}
\end{center}
\medskip

$^{\rm{a}}$Identification taken from C01 and B00. Only the primary period
			is given for RR01 stars.

$^{\rm{b}}$Type `lp' refers to long period variables of undetermined class. 

$^{\rm{c}}$Period found by our analysis, given to four decimal places.

$^{\rm{d}}$All periods are from CC01, with the exception of v154 \& v166, which are from C01, and v270 \& v271, which are from B00. Periods are rounded to four decimal places for comparison.

$^{\rm{e}}$Additional information about the star: pc--poor coverage of light curve;
		  m--merged with another star; n--significant scatter in light curve;
		nv--star does not appear to be variable;
		t--additional discussion of star in the main text; v--variable, but no clear period.

\end{minipage}
\end{table*}

\pagebreak

\begin{table*}
\begin{minipage}{120mm}
\caption{Data for new and suspected variable stars in the core of M3.}
\begin{center}
\begin{tabular}{lcccccc}
\hline\hline

ID & Type$^{\rm{a}}$ &  $\rm{R.A.(h:m:s)}^{\rm{b}}$ & $\rm{Dec.(deg:m:s)}^{\rm{b}}$ & Period$^{\rm{c}}$ & Comment$^{\rm{d}}$ \\
\hline
S1	&	RR1	&	13:42:06.5	&	28:20:59.2	&	0.3868	&	n, t	   	 \\
S2	&	RR1	&	13:42:17.7	&	28:21:06.2	&	0.2980	&	n, t  	 \\
S3	&	RR0	&	13:42:11.1	&	28:22:02.4	&	0.5162	&	n		 \\
S4	&	RR1	&	13:42:04.7	&	28:22:05.1	&	0.5075	&	t 	 	 \\
S5	&	RR1	&	13:42:09.9	&	28:22:32.1	&	0.3483	&	pc 		 \\
S6	&	RR0	&	13:42:10.7	&	28:22:17.9	&	0.6090	&	n, t 	       \\
S7	&	W Vir	&	13:42:12.8	&	28:22:49.6	&	1.3270	&	t 		 \\
S8	&	lp	&	13:42:10.7	&	28:22:58.1	&			&	pc, t, v	 \\
S9	&	RR0	&	13:42:04.9	&	28:22:06.0	&	0.4261	&	n, pc, t	 \\
S10	&	RR0	&	13:42:08.0	&	28:23:26.8	&	0.4778	&	n, t	  	 \\
S11	&	RR0	&	13:42:08.0	&	28:23:23.7	&	 		&	t, v	 	 \\
\hline

\end{tabular}
\end{center}
\medskip

$^{\rm{a}}$Type `lp' refers to long period variables of undetermined class. 

$^{\rm{b}}$Coordinates taken from our mapping using B00 data, as described in $\S 3.2.$

$^{\rm{c}}$Period found by our analysis, given to four decimal places.

$^{\rm{d}}$Additional information about the star: pc--poor coverage of light curve;
		  n--significant scatter in light curve;
		t--additional discussion of star in the main text; v--variable, but no clear period.

\end{minipage}
\end{table*}


\begin{thebibliography}{}
\bibitem[\protect\citeauthoryear{Alard and Lupton}{1998}]{al98}
  Alard, C. and Lupton, R., 1998, ApJ, 503, 325
\bibitem[\protect\citeauthoryear{Alard}{2000}]{a00}
  Alard, C., 2000, A\&AS, 144, 363
\bibitem[\protect\citeauthoryear{Bailey}{1913}]{b13}
  Bailey, S., 1913, Obs., 78, 1
\bibitem[\protect\citeauthoryear{Bakos et al.}{2000}]{b00}
  Bakos, G., Benko, J., Jurcsik, J., 2000, AcA, 50, 221 (B00)
\bibitem[\protect\citeauthoryear{Blazhko}{1907}]{b07}
  Blazhko, S., 1907, Astron. Nachr. 175, 325 
\bibitem[\protect\citeauthoryear{Carretta et al.}{1998}]{c98}
  Carretta, E., Cacciari, C., Ferraro, F., Fusi Pecci, F., Tessicini, G., 1998, MNRAS, 298, 1005 (C98)
\bibitem[\protect\citeauthoryear{Clement et al.}{2001}]{c01}
  Clement, C., Muzzin, A., Dufton, T. Ponnampalam, J., Wang, J., Burford, A., Richardson, T., Rosebery, J., Sawyer Hogg, H., 2001, AJ, 122, 2587 (C01)
\bibitem[\protect\citeauthoryear{Corwin and Carney}{2001}]{cc01}
  Corwin, M., Carney, B., 2001, AJ, 122, 3183 (CC01)
\bibitem[\protect\citeauthoryear{Evstigneeva et al.}{1994}]{e94}
  Evstigneeva, N., Samus', N., Tsvetkova, T., Shokin, Y., 1994 Pisma Astron. Zh., 20, 693
\bibitem[\protect\citeauthoryear{Guhathakurta et al.}{1994}]{g94}
  Guhathakurta, P., Yanny, B., Bahcall, J., Schneider, D., 1994, AJ, 108, 1786
\bibitem[\protect\citeauthoryear{Kholopov}{1977}]{k77}
  Kholopov, P., 1977, Perem. Zvezdy, 20, 313
\bibitem[\protect\citeauthoryear{Oosterhoff}{1939}]{o39}
  Oosterhoff, P., 1939, Obs., 62, 104
\bibitem[\protect\citeauthoryear{Sawyer Hogg}{1973}]{s73}
  Sawyer Hogg, H., 1973, P. D. D. O., 3, 1
\bibitem[\protect\citeauthoryear{Scargle}{1982}]{s82}
  Scargle, J., 1982, ApJ, 263, 835
\bibitem[\protect\citeauthoryear{Stellingwerf}{1978}]{s78}
  Stellingwerf, R., 1978, ApJ, 224, 953

\end{thebibliography}
\end{document}